\documentclass[useAMS]{mn2e}
\title{Detection and evolution of the  CO ($\Delta$$v$ = 2) emission in nova V2615 Ophiuchi (2007) }

\author[Das, Banerjee $\&$ Ashok]{R. K. Das,
\thanks{E-mail: rkdas@prl.res.in (RKD); orion@prl.res.in (DPKB);  ashok@prl.res.in (NMA)
}
D. P. K. Banerjee $\&$ N. M. Ashok  \\
Astronomy $\&$ Astrophysics Division, Physical Research Laboratory, Navrangpura, Ahmedabad 380009, India\\
}

\usepackage{graphicx}
\begin{document}

\pagerange{\pageref{firstpage}--\pageref{lastpage}} \pubyear{2008}

\maketitle

\label{firstpage}

\begin{abstract}

We present near-infrared (1 - 2.5 $\mu$m) spectroscopic and photometric results of Nova V2615 Ophiuchi which was discovered
in outburst in 2007 March. Our observations span a period of $\sim$ 80 days starting
from 2007 March 28 when the nova was at its maximum light.  The evolution of the spectra are shown from the initial  P-Cygni phase to an emission-line
phase and finally to a dust formation  stage. The characteristics of the $JHK$ spectra
are very similar to those observed in a nova outburst occurring on a carbon-oxygen white dwarf. We analyse an observed line at 2.088 $\mu$m and suggest it could be due to FeII excited by Lyman $\alpha$ fluorescence. The highlight of the observations is the
detection of the first overtone bands of carbon monoxide (CO) in the 2.29 - 2.40
$\mu$m region. The CO bands are modeled to  estimate the temperature and mass of the emitting CO gas and also to place limits on the $^{12}$C/$^{13}$C ratio. The CO bands are recorded
over several epochs thereby allowing a rare opportunity to
study its evolution  from a phase of constant strength through a stage when the CO is destroyed fairly rapidly. We compare the observed timescales involved in the evolution of the CO emission and find a good agreement with model predictions that investigate the
chemistry in a nova outflow during the early stages.
\end{abstract}


\section{Introduction}

V2615 Ophiuchi (Nova Ophiuchi 2007) was  discovered by Nishimura (2007) at a visual magnitude of 10.2  on
2007 March 19.812 UT. No star was visible at the position of the nova on a film
taken by Nishimura on 2007 March 17.82 UT and earlier survey films since
2005 (limiting mag 11.5-12).  The discovery report was supplemented by several other
observers from observations obtained close to and around the time of discovery.
These reports, in conjunction with the lack of a detection of the nova in images taken just a few days prior to discovery, imply that the nova was discovered   a few days before maximum light which was reached on 28 March.

The early low-resolution optical spectra (410-670 nm, resolution 1500) of V2615 Oph by Naito and
Narusawa (2007) on March 20.84 UT showed Balmer-lines having P-Cyg features and Fe II lines
(multiplets 42, 49, 74), suggesting that the nova is a "Fe II"-
type nova.  The FWHMs of the H$\alpha$, H$\beta$, and H$\gamma$ emissions
were 920, 810, and 790 km/s, and the displacement of the P-Cyg
absorptions from the Balmer emission peaks were 940, 820, and 830
km/s, respectively. Early optical spectra were also obtained by  Munari et al. (2007, 2008) on
Mar. 22.17 and 24.18 UT. On March 22.17 UT, the spectrum was characterized by
weak emission lines with P-Cygni profiles.  The
strongest emission lines were due to Fe II multiplets 27, 28, 37, 38, 42,
48, 49, 55, and 74; Si II multiplets 2, 3, 4 and 5; N I multiplet 3; Ca
II H and K; Na I D1 and D2; and O I 777.2-nm (the O I  line showed the second strongest emission component after H$\alpha$).   With respect to the peak of the H$\alpha$ emission, the
absorption terminal velocity was -1415 km/s, and the mean velocity -910km/s.
In comparison to the March 22.17 UT spectrum, the Mar. 24.18 UT spectra show large changes:
mainly a marked reduction in equivalent width of emission lines and reduction in
outflow velocity of absorption components of P-Cyg profiles.

Infrared observations of V2615 Oph have been made by Das et al. (2007) and Rudy et al.(2007) - some of the findings in Rudy et al. (2007) are discussed in details subsequently. The highlight of the Das et al. (2007) report was the detection of first-overtone CO  emission in the nova - similar emission has been recorded previously in only a few novae. In this paper, we present $JHK$ spectrophotometry
of V2615 Oph between March 28 and June 9. The spectroscopic results are fairly extensive as spectra have been obtained  on 15 occasions during this period.

\section{Observations}

The Near-Infrared $JHK$ spectrophotometric data of V2615 Oph presented here were acquired at the Mount Abu 1.2m telescope. The first observation was recorded on the night of  2007 March 28 while the nova was close to the optical maximum. The nova was then observed regularly until the arrival of the monsoon, when inclement weather brought our monitoring to an end. The instrument used was the Near Infrared Imager/Spectrometer with a 256$\times$256 HgCdTe NICMOS3 array. The observation logs for spectroscopy and photometry are presented in Tables 1 and 2 respectively.
Near-IR $JHK$ spectra were obtained at  similar dispersions of $\sim$ 9.75 {\AA}/pixel in each of the $J,H,K$ bands. In each band, spectra were recorded at different positions along the slit (slit width $\sim$1 arcsec) to provide for the background subtraction. To remove telluric features in the spectra of V2615 Oph, we obtained spectra of a nearby standard star (SAO 184301; spectral type A0V) at a similar airmass as the target.

The spectra were extracted using the APEXTRACT task in IRAF and the spectra were calibrated in wavelength using a combination of  OH sky lines and telluric lines in the extracted spectra. Following the standard reduction procedure, the nova spectra were then ratioed with the  spectra of the standard star from which the hydrogen Paschen and Brackett absorption lines had been manually removed. These ratioed spectra were multiplied by  a blackbody curve corresponding to the effective temperature of the standard star to yield the final spectra.

V2615 Oph was monitored photometrically in the JHK bands for 85 days after discovery. Photometric observations were performed under photometric sky conditions using the imaging mode of the NICMOS3 array. In each of the $J,H,K$ filters, several frames at 5 dithered positions offset typically by 20 arcsec, were obtained of both the nova and a selected standard star (2MASS J16232693 -2425291; $J$=7.340,$H$=6.027,$K$=5.464). Near-IR $JHK$ magnitudes were then derived using the IRAF aperture photometry task APPHOT. The derived $JHK$ magnitudes are given in Table 2, the typical errors associated with them lie in the range of 0.02 to 0.04 magnitudes.

\begin{table}
\caption{A log of the spectroscopic observations of V2615 Oph. The date of outburst
is taken to be 2007 March 19.812 UT}
\begin{tabular}{lcccccc}
\hline\
Date & Days        & &         & Integration time &   \\
2007 & since        & &        & (sec)            &    \\
(UT) & Outburst  & & \emph{J} & \emph{H}         & \emph{K}    \\
\hline
\hline

Mar. 28.928 & 09.116   & & 60      & 45               & 45 \\
Mar. 31.917 & 12.105   & & 60      & 60               & 60 \\
Apr. 02.901 & 14.089   & & 45      & 40               & 60 \\
Apr. 03.893 & 15.082   & & 60      & 40               & 45 \\
Apr. 05.904 & 17.092   & & 60      & 45               & 60 \\
Apr. 06.890 & 18.078   & & 60      & 60               & 60 \\
Apr. 07.915 & 19.103   & & 75      & 75               & 75 \\
Apr. 16.942 & 28.130   & & 90      & 90               & 90 \\
Apr. 18.888 & 30.076   & & 90      & 75               & 90 \\
Apr. 27.869 & 39.057   & & 90      & 60               & 90 \\
Apr. 30.830 & 42.018   & & 90      & 90               & 90 \\
May. 02.835 & 44.023   & & 90      & 90               & 90 \\
May. 06.832 & 48.020   & & 90      & 60               & 60 \\
May. 08.840 & 50.028   & & 90      & 90               & 90 \\
Jun. 09.846 & 82.034   & & 500     & 120              & 90 \\

\hline
\end{tabular}
\end{table}

\begin{table}
\caption{A log of the photometric observations of V2615 Oph. The date of outburst
is taken to be 2007 March 19.812 UT}

\begin{tabular}{lccccc}
\hline
Date & Days         & &             & Magnitudes&          \\
2007 & since        & &             &           &          \\
(UT) & Outburst     & & \emph{J}    & \emph{H}  & \emph{K} \\
\hline
\hline

Mar. 28.993	& 09.181 & &	6.26 &	5.83 &	5.39\\
Mar. 31.961	& 12.149 & &	6.94 &	6.51 &	5.90\\
Apr. 02.970	& 14.158 & &	6.49 &	6.07 &	5.54\\
Apr. 05.993	& 17.181 & &	6.54 &	6.14 &	5.64\\
Apr. 07.956	& 19.144 & &	6.93 &	6.57 &	6.18\\
Apr. 15.979	& 27.167 & &	7.36 &	7.07 &	6.53\\
Apr. 16.981	& 28.169 & &	7.22 &	6.97 &	6.46\\
Apr. 18.940	& 30.128 & &    7.30 &	6.96 &	6.62\\
Apr. 26.885	& 38.073 & &    7.76 &	7.53 &	7.09\\
Apr. 30.906	& 42.094 & &	7.98 &	7.68 &	7.18\\
May. 02.932	& 44.120 & &	8.14 &	7.84 &	7.07\\
May. 06.934	& 48.122 & &	8.13 &	7.72 &	6.79\\
May. 08.882	& 50.070 & &	7.94 &	7.32 &	6.26\\
Jun. 09.887	& 82.075 & &	9.31 &	7.62 &	5.34\\

\hline
\end{tabular}
\end{table}

\section{Results}
\subsection{Optical and near-infrared lightcurve}  The optical lightcurve of V2615 Oph is presented in
Figure 1. The object was discovered about 9 days before its maximum which was reached on 2007 March 28 at V = 8.52. On the whole, the object shows a steady post-maximum decline in brightness
of 0.05 mag/day for the first $\sim$ 80 days after maximum.
Subsequently there is a steep decline in the lightcurve due to dust formation - this aspect is
addressed shortly while discussing the near-IR light curve.
From the lightcurve we estimate $t$${_{\rm 2}}$ and $t$${_{\rm 3}}$ - the time for a drop of 2 and 3 magnitudes respectively in the visual brightness - to be
33 and 58 days respectively. The use of these values of $t$${_{\rm 2}}$ and $t$${_{\rm 3}}$ in various  MMRD relationships
(della Valle et al. (1990), Capaccioli et al. (1989) Cohen (1988) for $t$${_{\rm 2}}$; Schmidt (1957) for
$t$${_{\rm 3}}$) leads to closely similar values for the absolute visual magnitude - a mean
value of M${_{\rm v}}$ = 7.16 $\pm$ 0.12 is obtained. The extinction toward the object can be estimated in two ways. Rudy et al. (2007), using the strength of the OI lines, have determined the reddening prior to dust  formation to be  E(B-V) = 1.0 (or
A${_{\rm v}}$ = 3.1 magnitudes). This leads to an distance estimate of $D$ = 3.25 kpc to the object.
To show that this value of A${_{\rm v}}$ is reasonable, we use the extinction data of Marshall et al. (2006) which shows that  A${_{\rm Ks}}$ is constant at = 0.34-0.36
beyond 3 and upto 12 kpc in the direction of V2615 Oph.
Thus, assuming A${_{\rm v}}$/A${_{\rm Ks}}$ $\sim$ 11 (Koornneef 1983), a maximum value of  A${_{\rm v}}$ = 3.85 is suggested toward V2615 Oph thereby leading to a lower limit on the distance of $D$ = 2.3 kpc. For the present work we adopt the distance estimate derived from the Rudy et al (2007) findings viz. $D$ = 3.2 kpc.

Our IR observations, begun close to optical maximum, show a steady decline for the first 42 days. After this each of the $J$,$H$,$K$ magnitudes show
an increasing trend possibly indicating the onset of dust formation. However, no corresponding
sharp decline in the optical light curve is seen at this stage implying that any dust, that
may have formed, is likely to be optically thin.
Rudy et al. (2007) have  reported 0.4 to 2.5 $\mu$m spectroscopy of V2615 Oph
on May 7 2007 and 0.8 to
5.5 $\mu$m spectroscopy on 2007 May 31. They find that the nova evolved dramatically
between the two observations (i.e between May 7 and May 31) due to
the formation of dust.  The reddening derived from the OI lines
increased from E(B-V)=1.0 to 1.3 in their data. Our
last observations on 9 June 2007 show a significant infrared excess due to emission by the newly formed dust.

\begin{figure}
\centering
\includegraphics[bb= 5 3 478 403, width=3.0in,height=3.0in,clip]{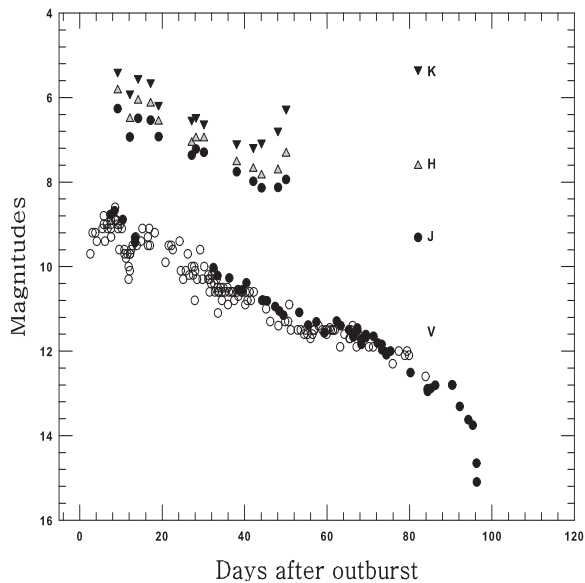}
\caption[]{ The $V$ band lightcurve is shown at the bottom (photoelectric/CCD measurements shown by filled  circles ; visual estimates by open circles) based on data
 from IAU circulars, AAVSO (American Association of Variable Star Observers) and AFOEV databases (Association Francaise des Observateurs d'Etoiles Variables, France). At the top are shown the
near-IR $J$, $H$ and $K$ band light curves based on the present observations ($J$ : filled  circles, $H$ : gray triangles - up;  $K$ : black triangles - down.)}
\label{fig1}
\end{figure}


\begin{figure}
\centering
\includegraphics[bb= 2 3 263 405, width=3.0in,height=5.0in,clip]{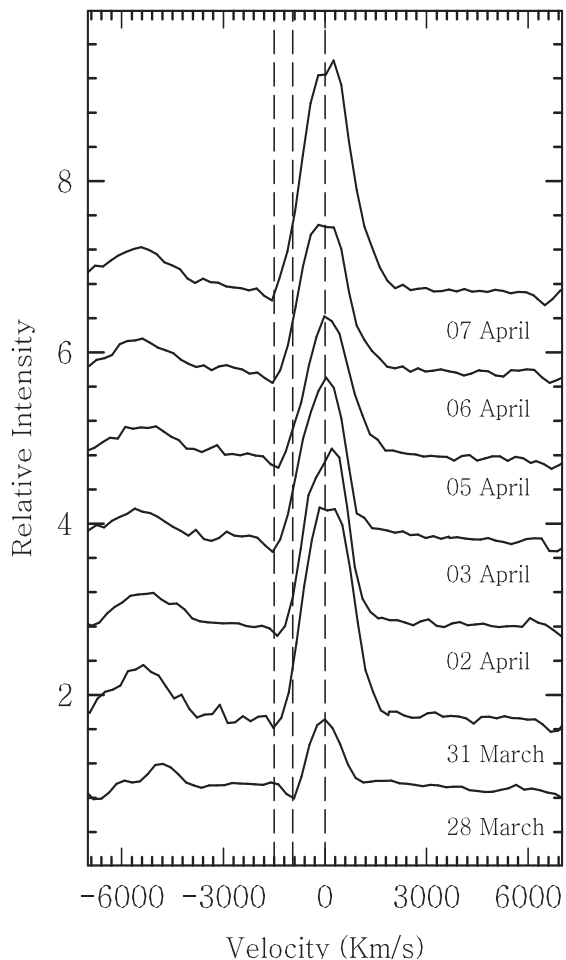}
\caption[]{ The  Paschen $\beta$ 1.2818 $\mu$m line is presented on different
days to show the evolution of the profile. The two dotted lines,
at left of the line center, are positioned at -950 and -1500 km/s respectively to
indicate the change in position of  the P-Cygni absorption minimum with respect to the emission peak (further details are in the text of Section 3.2)}
\label{fig1}
\end{figure}


\subsection{General characteristics of the $JHK$ spectra}
The lines in the earliest spectrum of 28 March, when the nova was at maximum light, show P-Cygni structure with  a prominent absorption component. By following the
Pa $\beta$ line for example,  it is seen that  the P-Cygni absorption component  persists for at least 10 days (upto 7 April)  indicating
that mass loss extends over a prolonged period. The minimum of the absorption component
is found to be displaced from the emission peak by $\sim$ 950 km/s on 28 March but this value changes closer to $\sim$ 1400-1500 km/s for 31 March and other subsequent dates upto 7 April. Variations, of a similar nature, in the outflow velocities were also noted by Munari et al (2007) on 22 and 24 March - when the nova was on its rise to maximum light. Clearly, the mass loss process and its kinematics during the early stages is not uniform.

Mosaics of the
the $J$, $H$ and $K$ band spectra are presented in Figures 2, 3, and 4 respectively.
The early spectrum of V2615 Oph is typical of a carbon-oxygen nova (CO nova) - examples of which are  V2274 Cyg (Rudy et al. 2003) , V1419 Aql (Lynch et al. 1995) and V1280 Sco (Das et al. 2008). In addition to lines of H, He, N and O, these novae show strong lines of carbon e.g  in the $J$ band in the 1.16 to 1.18 $\mu$m wavelength region. In contrast, such carbon lines are weak  in the spectra of ONeMg novae e.g. in the novae V597 Pup and V2491 Cyg (Naik et al. 2009). The IR-based classification of V2615 Oph is consistent with  its optical classification as a FeII type nova (Naito $\&$ Narusawa, 2007). Fe II novae are believed to be associated  with explosions on CO white dwarfs (Williams 1992). We have presented the line  identification  in Table 3 but do not show them marked on the spectra as a separate figure. However, since the lines are very similar to those seen in V1280 Sco,
 the line-identification figure for V1280 Sco could be referred to which contains greater details (Das et al. 2008). The most prominent lines in the $JHK$ spectra of V2615 Oph are the Paschen and Brackett hydrogen recombination lines; the Lyman $\beta$ and continuum fluoresced OI lines at
1.1287 and 1.3164 $\mu$m; the HeI lines at   1.0830 $\mu$m and 2.0585.
Strong carbon lines are seen in the $J$ band also in the $H$ band redwards
of Br 11. Among subtle features,  the presence of a C I line at 1.6 ${\rm{\mu}}$m, which could  be mistaken as just another member of the $H$ band Brackett series lines, should not be missed. The region between 1.2 to 1.275 ${\rm{\mu}}$m contains the complex blend of a  large number of   NI and C I lines that
are  often seen in the early spectra of CO novae. The presence of several lines of NaI  and MgI ( the prominent ones being at 1.1404, 1.5040, 1.5749, 2.2056
and 2.2084 $\mu$m ) are worth noting
as it may be inferred, from their presence, that dust will form in a nova.
During the analysis of V1280 Sco, one of the conclusions  that emerged, was that the  lines of NaI particularly,  and also of MgI,  are associated with low excitation and ionization
conditions (Das et al. 2008). Such conditions    necessarily imply the  existence  of a cool zone which  is conducive for dust formation.  This was observationally
corroborated  in the case of several novae, in which   these  lines were detected, and which proceeded to form dust (e.g.  V2274 Cyg (Rudy et al. 2003), V1419 Aql (Lynch. et al. 1995),  NQ Vul  (Ferland et al 1979), V705 Cas (Evans et al 1996),
 V842 Cen (Wichmann et al. 1991), V1280 Sco (Das et al. 2008); for a more detailed discussion see Das et al. 2008).  The presence of these lines in V2615 Oph, and the dust formation witnessed subsequently,  is consistent with this scenario. The formation
of dust is clearly reflected in the last spectrum of 9 June. Each of the spectra
in the individual $J$, $H$ and $K$ bands shows the continuum level increasing to longer wavelengths showing the development of an infrared excess associated with dust.

The most interesting spectral features in V2615 Oph are the first overtone CO bands
in the $K$ band which is discussed in coming sections. The other atomic lines in V2615 Oph are similar to those seen in carbon-oxygen novae and which were discussed   in our earlier work on V1280 Sco - hence they are not discussed
further here. There is, however, an unidentified line at 2.088 $\mu$m which is discussed below.

\begin{figure*}
\centering
\includegraphics[bb=2 1 684 757,width=7.0in,height=7.5in]{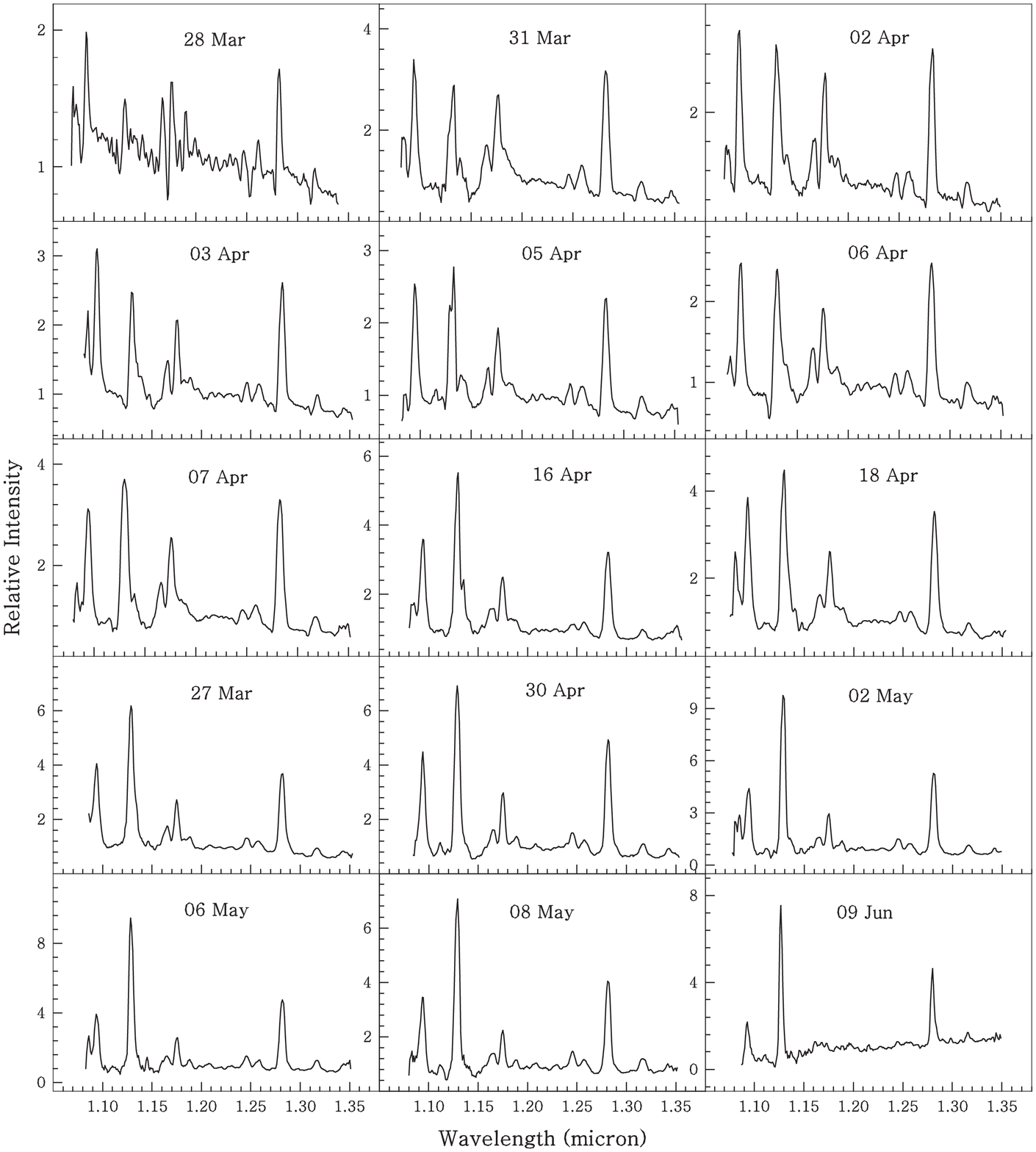}
\caption[]{ The $J$ band spectra of V2615 Oph  on different days with the flux normalized to unity at 1.25 ${\rm{\mu}}$m.
}
\label{fig2}
\end{figure*}

\begin{figure*}
\centering
\includegraphics[bb=2 1 691 759,width=7.0in,height=7.5in]{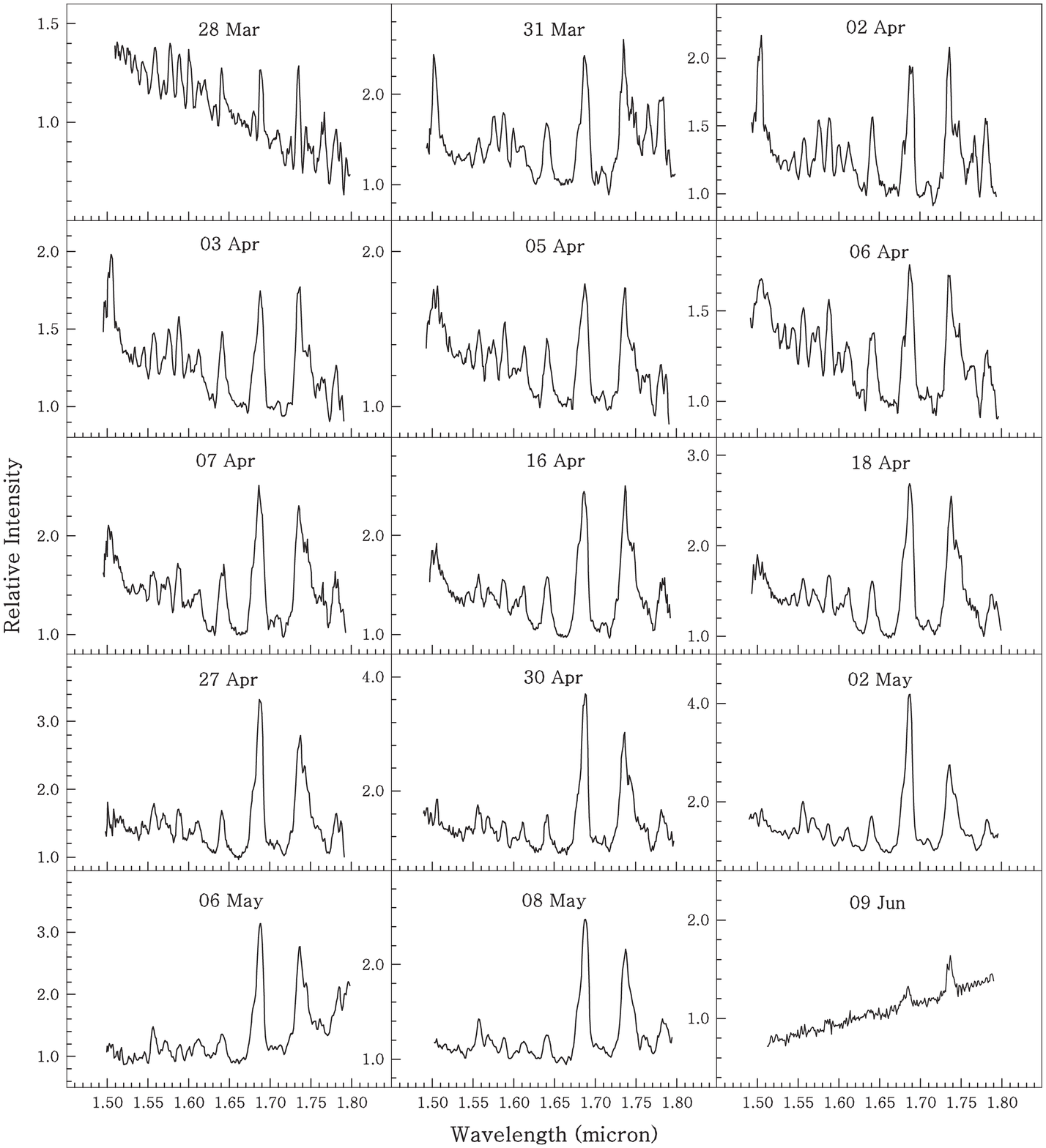}
\caption[]{ The $H$ band spectra of V2615 Oph on different days with the flux normalized to unity at 1.65 ${\rm{\mu}}$m.}
\label{fig2}
\end{figure*}

\begin{figure*}
\centering
\includegraphics[bb=1 0 725 785,width=7.3 in,height=7.5in,clip]{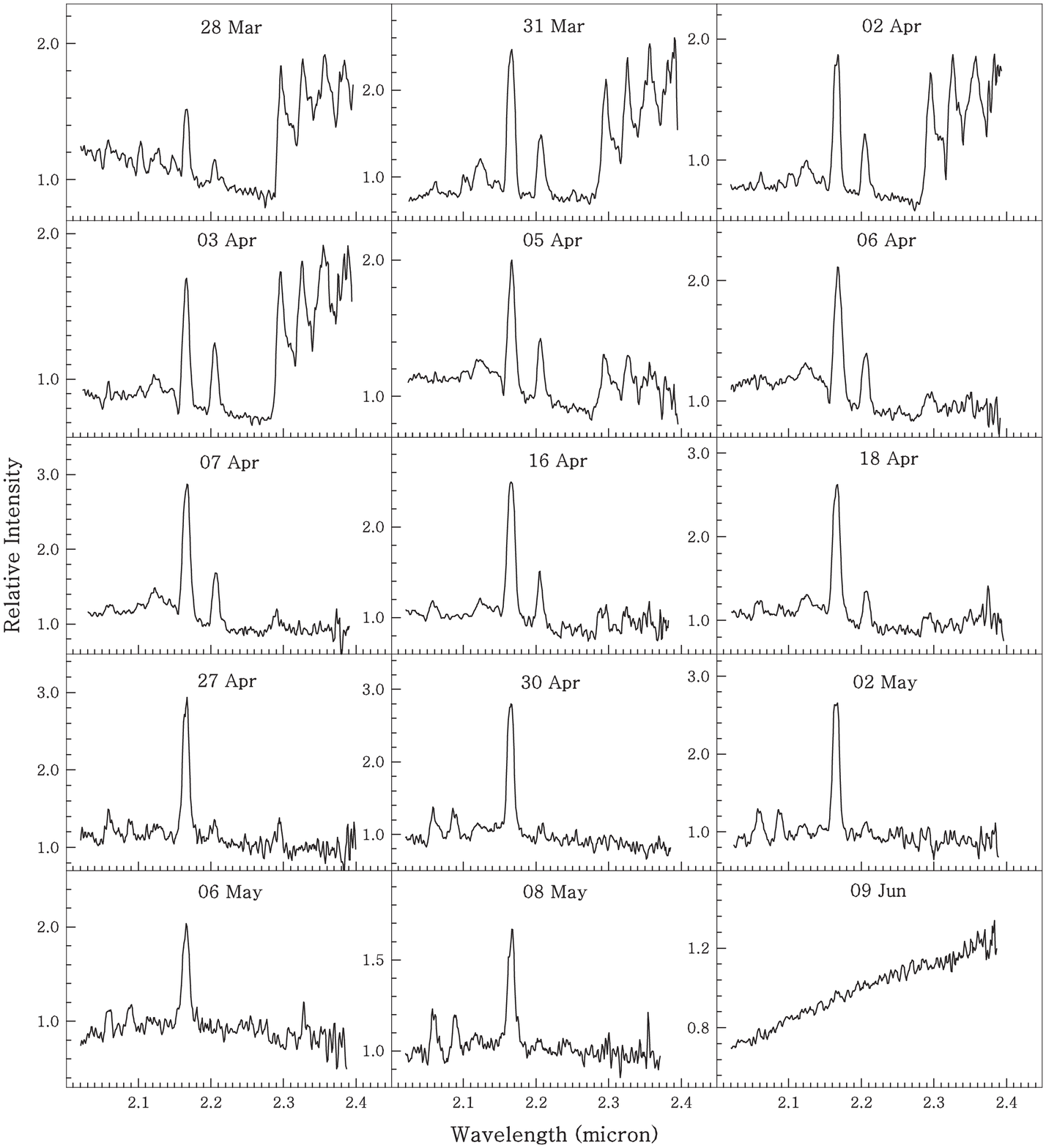}
\caption[]{ The $K$ band spectra of V2615 Oph on different days with the flux normalized to unity at 2.2 ${\rm{\mu}}$m.
}
\label{fig2}
\end{figure*}


\begin{table}
\caption{List of observed lines in the $JHK$ spectra}\
\begin{tabular}{lll}
\hline\\
Wavelength         	& Species          	& Other contributing  \\
(${\rm{\mu}}$m)    	&                  	& lines $\&$ remarks  \\
\hline
\hline \\
1.0830             	& He \,{\sc i}      & 				    \\
1.0938   	   	    & Pa $\gamma$     	&        			\\
1.1116              & u.i.              &                   \\
1.1126   		    & Fe \,{\sc ii}     &               	\\
1.1287   		    & O \,{\sc i}      	& 				    \\
1.1330   		    & C \,{\sc i}      	& 				    \\
1.1381,1.1404       & Na \,{\sc i}      &                   \\
1.1600-1.1674 		& C \,{\sc i}      	& strongest lines at 	\\
                    &                 	& 1.1653, 1.1659,1.16696\\
1.1748-1.1800 		& C \,{\sc i}      	& strongest lines at \\
                    &                 	&1.1748, 1.1753, 1.1755		\\
1.1828			    & Mg \,{\sc i}      &                           \\
1.1880       		& C \,{\sc i}       &blended with C I at 1.1896 \\
1.2074,1.2095   	& N \,{\sc i}       &blended with C I at 1.2088    \\
1.2187,1.2204  		& N \,{\sc i}    &                          \\
1.2249,1.2264  		& C \,{\sc i}    & 						\\
1.2329   		    & N \, {\sc i}   &  \\
1.2382   		    & N \,{\sc i}    &  \\
1.2461,1.2469  		& N \,{\sc i}    &                \\
1.2562,1.2569  		& C \, {\sc i}   &    \\
1.2818   		    & Pa $\beta$       	& 	  \\
1.3164   		    & O \,{\sc i}		&     \\
1.3465              & N \,{\sc i}       &     \\
1.5040              & Mg \,{\sc i}      & \\
1.5184   		    & Br 20             &                 	\\
1.5256   		    & Br 19             &   				\\
1.5341   		    & Br 18            	&  				    \\
1.5439   	   	    & Br 17            	&  				    \\
1.5557   		    & Br 16            	&   				\\
1.5685  		    & Br 15             &    				\\
1.5749              & Mg \,{\sc i}      &         \\
1.5881   		    & Br 14             &    \\
1.6005  		    & C \,{\sc i}       & 				\\
1.6109    		    & Br 13    		    &  				\\
1.6407  		    & Br 12       		&  \\
1.6806   		    & Br 11    		    &  				\\
1.6890   		    & C \,{\sc i}   	&  				\\
1.7109              & Mg \,{\sc i}      &  				\\
1.7200-1.7900  		& C \,{\sc i}       & Several  C \,{\sc i} lines	\\
1.7362  		    & Br 10      		& 	\\
2.0585 			    & He \,{\sc i}      & blended with u.i. 2.0620 \\
2.0888			    & u.i.              & Fe \,{\sc ii} ? (See section 3.3) \\
2.1023 			    & C \,{\sc i} 		&  \\
2.1156-2.1295  		& C \,{\sc i}     	& blend of several  C \,{\sc i} lines		\\
2.1452   		    & Na \,{\sc i}   	& 		\\
2.1655   		    & Br $\gamma$     	&  				\\
2.2056,2.2084	    & Na \,{\sc i}      & 				\\
2.2156-2.2167		& C \,{\sc i}  		&  		\\
2.29-2.40           & CO               & $\Delta$v=2 bands             \\
2.2906   		    & C \,{\sc i}      	& 				\\
2.3130              & C \,{\sc i}       &               \\
2.3348,2.3379       & Na \,{\sc i}      &                   \\

\hline
\end{tabular}
\end{table}

\subsection{An unidentified line at 2.0888 $\mu$m:  possibly a FeII line excited by Lyman $\alpha$  fluorescence?}

We note the presence of an emission line at $\sim$ 2.0890 $\mu$m in the $K$ band
that is seen fairly prominently in the spectra from 27 April onwards.
We propose that this is a FeII line and additionally investigate whether  this   line could be  excited by
Lyman $\alpha$ (Ly$\alpha$) fluorescence. In the near-infrared, there are a few FeII lines seen in the spectra of novae, which are believed to be primarily excited by  Lyman $\alpha$ and Lyman continuum fluorescence. Among these are  the  so-called "one micron Fe II lines" at  0.9997, 1.0171, 1.0490, 1.0501, 1.0863, and 1.1126 $\mu$m  seen in several novae (Rudy et al. 1991, 2000).   In addition, two other Fe II lines at 1.6872 and 1.7414 $\mu$m in the $H$ band, are also proposed to be pumped by the same mechanism (Banerjee et al. 2009). The $H$ band lines are prominently detected in the 2006 outburst of recurrent nova RS Oph (Banerjee et al. 2009),  in the slow nova V2540 Ophiuchi  (Rudy et al 2002) and possibly also in the recurrent nova  CI Aquila
(Lynch et al. 2004). These $H$ band lines could be present in the spectra of other novae too, but  have evaded  detection because of blending with the  Br 11 (1.6806 $\mu$m) and CI lines at 1.6890 and 1.7449
$\mu$m which lie close by. However these lines are  well resolved  in RS Ophiuchi (Banerjee et al. 2009), especially during the later stages of its outburst when all line widths in RS Oph become  small - due to deceleration of the shocked, emitting gas - and blending effects are thereby minimal.

The  excitation mechanism of the Fe II
lines appears to be  a three step process (Johansson $\&$ Jordan 1984; Bautista et al. 2004 and references therein, Banerjee et al. 2009). FeII is first excited from low-lying levels by Ly$\alpha$ or Lyc to a high energy level (typically 11 to 13 eV above ground state) which decays, in the second step, to feed the upper level associated with the observed FeII line  (the decay of this upper level, in the third step, leads to the line proper).  The 2.0888 $\mu$m line results from the decay of the  $3d^6(^3_2F)4s$ $c^4\!F$ term at $\sim$ 6.209 eV above the ground state.
It so happens that this term can be fed by not just
one, but in fact by several high lying levels - each of these high lying
levels capable of being  pumped by Ly $\alpha$ photons. As illustrative examples,
it is  noted that  photons at 1213.738, 1214.067, 1216.239, 1216.272 and
1217.152 {\AA}, which
are reasonably coincident with the Ly $\alpha$ line center at 1215.671$\AA$, can excite
transitions from the low lying (5D) 4s $a^4\!D$ term  in Fe II (at 1.076 eV above ground
state) to the  higher excited levels  ((5D) 4p 4F, (5D) 4p 4D and (3P) 4p 4P
respectively at around 11.3 eV ). Since HI  lines in novae (Ly $\alpha$
included) are broad with widths extending upto a few thousands of
km/s (1 Angstrom corresponds to about 250 km/s at the Ly $\alpha$ wavelength),  these photons could contribute to the flourescence process, even though they are not coincident with the Ly $\alpha$ line centre. These  higher
levels at 11.3 eV can then decay via ultraviolet photons  to the upper level of the 2.0888 $\mu$m transition.   The
Kurucz atomic data {\footnote{http://cfa-www.harvard.edu/amp/ampdata/kurucz23/sekur.html}} on which the present analysis is based, indicates  there are
additional Ly$\alpha$ fluorescing candidate lines (apart from the five discussed here),
all within a few Angstroms of the  Ly $\alpha$ line center  - that could also contribute to the Ly $\alpha$  fluorescence process. Therefore, the 2.0888 $\mu$m line could be excited by Ly alpha flourescence, if its identification with Fe II is correct.

A few cautionary words regarding the identification of the 2.0888 $\mu$m with FeII may be in order.
In case the Fe II identification is correct, then a few other Fe II lines - as mentioned earlier -  could also be expected viz. lines at   0.9997, 1.0171, 1.0490, 1.0501, 1.0863,  1.1126, 1.6872 and 1.7414 $\mu$m. Unfortunately the first four of these lines are not covered in our spectra while it is difficult to make any definitive conclusion about the 1.0863 $\mu$m line which is in a region of low signal since it is at the edge of our spectral window. However the 1.1126 $\mu$m line is seen. It is also difficult to draw a firm conclusion whether the 1.6872 and 1.7414 $\mu$m lines are present. Unfortunately both these lines occur at positions close to   strong CI and HI lines  (Table 3)
which could lead to line blending. In essence, further detections of the 2.0888  $\mu$m line in other novae is desirable to enable a secure identification.

\subsection{Modeling the CO emission}

The CO emission is modeled by assuming the gas to be  in thermal equilibrium characterized by the same rotational and vibrational temperature. The populations of the different levels can then be determined from the Boltzmann distribution since  the energy  of  individual rovibrational levels is known (we adopt a similar model as Spyromilio et al. 1988).
In our calculations, we have used   values for the
rotational and vibrational constants  of $^{12}$C$^{16}$O, $^{13}$C$^{16}$O from the NIST database
{\footnote{http://physics.nist.gov/PhysRefData/MolSpec}} except for the vibrational
constants for $^{13}$CO which are adopted from Benedict et al. (1962). We have not considered other isotopic species like $^{12}$C$^{17}$O, $^{14}$C$^{16}$O etc. since they are expected to have low abundances.

The  line luminosity $E$ of each rovibrational
transition  can  be calculated from knowing
the population of the upper level involved in the transition, the associated transition probability for the line  (Goorvitch 1994) and the
photon energy ($h$$\nu$) associated with the transition. To enable a comparison of the model value with the observed data, the line luminosity $E$ calculated above, which is in units of ergs/sec,   needs to be converted into units of the observed flux (we use units of ergs/s/$\rm cm^{2}$/$\mu$m). This conversion is achieved by  first  dividing $E$ by 4$\pi$$D^2$ (where $D$ is the distance to the source) and subsequently scaling $E$ to a unit of wavelength. In implementing the last step, it is assumed that each rotational line  is gaussian in shape.  Then, the peak intensity of such a gaussian (plotted with its ordinate units in ergs/s/$\rm cm^{2}$/$\mu$m; abscissa in units of $\mu$m) can be determined analytically by ensuring that the integrated area under this gaussian matches the known quantity $E$/4$\pi$$D^2$.
The gaussian line profiles of all the rotational lines are thus generated and co-added together to generate the resulting envelope of the CO emission. To this envelope we add the appropriate continuum, the level of which is determined from broad band photometry of Table 2,  to provide a model CO emission spectrum for a particular day. Such model spectra matching the observed data of 28 and 31 March; and 2 and 3 April are shown in Figure 5. The parameters that are fed  as inputs for the modeling are
the total mass of the CO gas ($M_{CO}$), the $^{12}$C/$^{13}$C   ratio (designated as $\alpha$) and the gas temperature ($T_{CO}$).
Once $M_{CO}$ and $\alpha$ are chosen, the total number of  $^{12}$CO and $^{13}$CO  molecules are fixed - thus the level populations  in thermal equilibrium at a temperature $T_{CO}$ can be calculated and the emerging spectrum determined. The estimated model CO flux is  therefore an absolute quantity but it could however be subject to certain errors.
One of these is the value of the distance $D$ to the source  - the model CO flux scales as $D^{2}$.

Different combinations of the parameters $M_{CO}$, $\alpha$ and $T_{CO}$ were tried to find the best fit between model spectra and  the observed data. The characteristics of the CO emission remains fairly constant between 28 March and 3 April. The profiles of Figure 6 are reasonably well modeled with closely similar values of $T_{CO}$ and $M_{CO}$ in the range of  4000 - 4300K and  2.75x$10^{-8}$  to  3.25x$10^{-8}$ $M_{\odot}$ respectively. Some of the difficulties encountered in the modeling, and also possible sources of errors involved therein, are as follows. It is noted that changes brought about in the spectra by either changing
 $T_{CO}$ or changing  $M_{CO}$ can be similar  in the following  respect. Increasing either of these quantities enhances  the absolute level of the CO emission.
However, the contributions of these two parameters can be distinguished by the fact that  increasing $M_{CO}$ just scales up the overall level of the CO emission level.
On the other hand, increasing $T_{CO}$ not only increases the CO flux but additionally changes the intensities of the different vibrational bands, relative to each other, within the first overtone. Therefore, vibrational bands from v$\geq$5, would help the analysis, but they are located in regions of poor atmospheric transmission, beyond the spectral coverage and the v=5-3 band is barely covered.

While making the fits we note that the CO bands are likely contaminated by CI lines at 2.2906 and
2.2310 $\mu$m and from NaI lines at 2.3348 and 2.3379 $\mu$m. The position of these lines are marked in the bottom panel of Figure 6 and there are discernible structures at these positions,  in most of the profiles, indicating these lines are present. The 2.29 $\mu$m ($v$ = 2-0) band  would appear to be  affected, especially by the CI 2.2906 line which can be significant in strength in carbon-oxygen novae (Rudy et al. 2003; Das et al. 2008) and whose presence is possibly responsible for lack of agreement between model and observed data in this region. In view of the above, we have relied more on the fits to the higher bands ( the $v$ = 3-1 and 4-2 bands)
while estimating the CO parameters.  We have also assumed the CO emission to be optically thin. For the assumptions and constraints outlined above, the formal fits of Figure 6 yield estimates of $T_{CO}$ of 4100, 4000, 4100 and 4300K (with an error
of $\pm$ 400K) for 28 March, 31 March, 2 April and 3 April. We find that a
constant  CO mass of  3.0x$10^{-8}$ $M_{\odot}$, or at most a marginal variation
 of $M_{CO}$ between 2.75x$10^{-8}$  to
3.25x$10^{-8}$ $M_{\odot}$, can account for the observed profiles on the four days.  The fits shown in all panels of Figure 6 are made for the same
mass of  3.0x$10^{-8}$ $M_{\odot}$ and the implications of a fairly constant mass
is discussed shortly.

As in earlier studies on the $^{12}$C/$^{13}$C ratio in a few novae, discussed in the
following sub-section, we are also able to place only a lower limit on this ratio viz.  $^{12}$C/$^{13}$C is greater than 2. If this ratio is decreased below 2, the $^{13}$CO contribution becomes rather prominent i.e. the $^{13}$CO bandheads begin to prominently appear in the  synthetic spectra resulting in poor model fits. The determined lower limit for the $^{12}$C/$^{13}$C ratio may be compared with that expected from theoretical calculations. For  carbon-oxygen novae like V2615 Oph, the expected $^{12}$C/$^{13}$C ratio in the ejecta has been computed (e.g. Jose and Hernanz (1998); Starrfield et al. (1997)) and shown to be dependent on the white dwarf mass among other parameters. Different models by Jose and Hernanz (1998) show that this ratio is approximately constrained between 0.3 to 0.65 for a white dwarf mass between 0.8 to 1.15 $M_{\odot}$; Starrfield et al. (1997) find the $^{12}$C/$^{13}$C ratio to decrease from 2.4 to 0.84 as the white dwarf mass increases form 0.6 to 1.25 $M_{\odot}$. However, our observations and modeling seem  to indicate that $^{13}$C is possibly not synthesized to the high levels predicted by the theoretical models.

To calculate the CO column density, we assume that the CO is uniformly spread in a  shell of thickness $\Delta$R and volume 4$\pi$$R^2$$\Delta$R. The radius R of the shell is estimated kinematically knowing
the time elapsed since outburst and the velocity of the shell. A value of
$\sim$ 1450 km/s is adopted for the shell velocity based on the P-Cygni terminal velocity reported
by Munari et al. (2007) and similar values of the FWHM of the near-IR lines as found in this work. The column density, which is proportional to ($M_{CO}$$\Delta$R)/(4$\pi$$R^2$$\Delta$R), is then independent of
the thickness of the shell. In this manner we determine the CO column densities to be 7.5x$10^{18}$, 4.1x$10^{18}$, 3.8x$10^{18}$ and 2.9x$10^{18}$ $\rm cm^{-2}$ on 28 March, 31 March, 2 April and 3 April respectively.

It maybe useful to present similar results in other known novae in which the first
overtone of CO has been detected. These are V2274 Cyg (Rudy et al. 2003), NQ Vul (Ferland et al. 1979), V842 Cen ( Hyland $\&$ Mcgregor 1989, Wichmann et al. 1990, 1991), V705 Cas (Evans et al. 1996) and  V1419
Aql (Lynch et al. 1995). The CO mass $M_{CO}$ was determined to be  $10^{25.5 \pm 0.5}$ gms (i.e. 1.6x$10^{-8}$ $M_{\odot}$) in NQ Vul by Ferland et al. (1979) from observations 19 days after  outburst or 20 days before the
large visual fading associated with dust formation. A limit of $^{12}$C/$^{13}$C greater than 3.0  and a temperature $T_{CO}$ = 3500 $\pm$ 750K were determined for the object. In V2274 Cyg, Rudy et al. (2003)
 from observations 17 days after discovery, determined $M_{CO}$ = 8x$10^{-9}$ $M_{\odot}$, $^{13}$C/$^{12}$C $\ge$
0.83 $\pm$ 0.3 and $T_{CO}$ = 2500K. In nova V705 Cas, observations by Evans et al. (1996) taken at two epochs i.e. one day before maximum light and 26.5 days after
maximum light yielded estimates for $M_{CO}$  of (2.8 $\pm$ 0.2)x$10^{-10}$ $M_{\odot}$ and  (3.8 $\pm$ 0.2)x$10^{-10}$ $M_{\odot}$ respectively and $T_{CO}$ equal to
4300 $\pm$ 300K $\&$ 4500 $\pm$ 300K respectively. They estimate
the $^{12}$C/$^{13}$C ratio to be $\ge$ 5 and the CO column density to be
2x$10^{17}$ $\rm cm^{-2}$ for $M_{CO}$ equal
to 1.0x$10^{-10}$ $M_{\odot}$. Apart from the above sources, the $^{12}$C/$^{13}$C ratio has also been estimated in two other novae. In DQ Her this ratio was found to be $\ge$ 1.5  by Sneden and Lambert (1975); in V842 Cen observations between days 29 and 45 after
outburst by Wichmann et al. (1991) show the $^{12}$C/$^{13}$C ratio to be 2.9 $\pm$ 0.4.

\begin{figure}
\centering
\includegraphics[bb=1 0 295 613,width=3.5in,height=8.0in,clip]{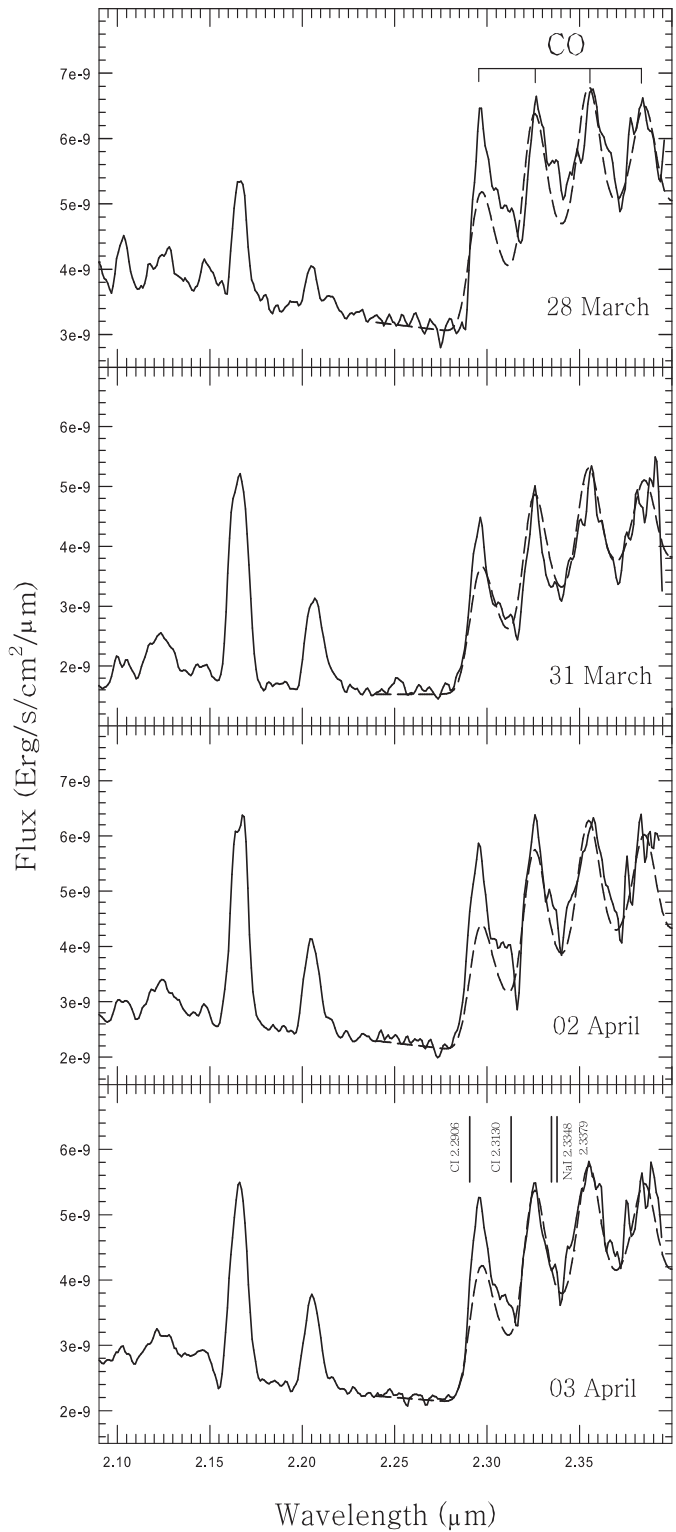}
\caption[]{ Model fits (dashed lines) to the observed first overtone CO bands
in V2615 Oph. The fits  are made for a constant  CO mass
of  3.0x$10^{-8}$ $M_{\odot}$ on all  days while the temperature of the gas
$T_{CO}$ is estimated to be  4100, 4000, 4100 and 4300K (with an error
of $\pm$ 400K) for 28 $\&$ 31 March and  2 $\&$ 3 April respectively. The bottom panel
shows the position of certain CI and NaI lines that complicate the modeling.
The other prominent lines seen in the spectra are Br $\gamma$ at
2.1655 $\mu$m; NaI 2.2056, 2.2084 $\mu$m blended with weaker emission from CI 2.2156, 2.2167 $\mu$m
lines and other CI lines between 2.1 and 2.14 $\mu$m. The position of the $^{12}$C$^{16}$O bandheads are shown in the top panel.
}
\label{fig2}
\end{figure}

\subsection{Evolution of the CO emission}
The detection of CO over a significant duration of time in the present observations,
presents an opportunity - not available before - to study  the  formation and destruction of CO during a  nova outburst. The earliest theoretical studies of the chemistry of novae were done by Rawlings (1988)   in the form of pseudo-equilibrium chemical models of the pre-dust-formation epoch. These models, which were developed with the main aim of explaining the observed presence of CO in novae, found that the outer parts  of the ejecta have to be substantially more dense and less ionized than the bulk of the wind for substantial molecule formation to occur. For this to occur carbon has to be neutral. In a neutral carbon region, the carbon ionization continuum, which extends to less
than  1102 $\AA$, shields several molecular species against the dissociative UV flux from the central star. A more refined  model for molecule formation in the nova outflow in the early stages  is presented in Pontefract and Rawlings (2004; hereafter PR) - we try to correlate the present observational findings with the results in this work. The PR studies are a major extension of their earlier models with only one major  qualitative point of departure viz.  neutral-neutral reactions are now shown to be more important than photoreactions in determining the nova chemistry.

A significant result in PR is the prediction of the evolution of the
fractional CO abundance with time. Two models are considered -  model A considers oxygen rich ejecta and model B considers carbon rich ejecta. Figures 1 and 2 of PR show the evolution of the fractional abundance of different molecules and radicals, including CO,  with time. It is seen that in both models,  the CO abundance remains constant upto about 2 weeks after
outburst ($\sim$ 12 days in case A and $\sim$ 15 days in case B). This behaviour,
and the length of its duration, seems to be generic to the models. During this phase the CO is saturated - that is to say, all the available oxygen or carbon, whichever has the lower abundance, is completely incorporated into forming
CO. After this there is a sharp decline in the CO abundance as CO is destroyed mostly by
reactions with N and $\rm N^+$. During this stage, from Figures 1 and 2 of PR, we estimate an approximate decrease in CO  by a factor of 1000 in $\sim$ 27days for model A and a decrease by a factor of 100 in $\sim$ 16 days for model B. The present data
allows to check on these two vital aspects of the model predictions viz. the
existence of a short-lived saturated phase followed by a phase involving rapid
destruction of CO.

The present observations and modeling show that the CO mass was  constant between 28 March and 3 April i.e. for a period of 7 days. This puts a lower limit on the duration of the saturated phase  since our observations  commenced on 28 March, nearly 8-9 days after the beginning of the outburst on Mar. 19.812 UT. It is possible that the CO emission was present, and at  similar levels, between March 19 and March 28 also. After all, CO has been seen very early after commencement of the eruption as in the case of V705 Cas (Evans et al. 1996). If that is the case, then the upper limit on the saturated-phase timescale would be around 15 days. Thus the evidence indicates that a phase does exist when the CO mass is constant and whose duration $t_s$ is constrained within 7 $\le$ $t_s$ $\le$ 16 days.
This observational finding conforms  well with the predicted timescales of $\sim$ 12-15 days from the
PR model calculations

Between 3 April and 5 April there is a sudden decrease in the CO strength. It is difficult to meaningfully model the 5 April data, and hence derive the CO parameters, because the relative contribution of the CI 2.2906 $\mu$m line is  considerable at this stage. Also the signal-to-noise ratio in the region around the CO emission region begins to get low now. However, if we assume that $T_{CO}$ has not changed drastically between 3 and 5 April, then from the diminished CO flux on the latter date, we estimate that $M_{CO}$ has decreased by a factor of 3. Beyond this date,  the decrease in the CO emission continues to take place, with a possible restrengthening
again on 16 April, to finally drop below measurable limits by 27-30 April. But the data beyond 5 April is of inadequate quality to  make a quantitative assessment of the strength of the CO emission beyond this stage (or indeed whether CO is even present on some of the days; the presence of the CI lines complicates the assessment further). However, a quantitative assessment can certainly be made with a good degree of  surety - there is a phase of rapid reduction in the CO emission after 3 April.
The initial decrease is indeed rapid - the spectra over a 4 day gap, between 3 April
 and 7 April, have only to be compared to establish this. This quantitative behaviour i.e. the rapid destruction of CO following the saturated phase  is again
largely consistent with the predictions  of the theoretical model (PR; 2004).

It is also possible to estimate the CO:C ratio, though with substantial uncertainties, at different stages of the CO evolution and compare it with model predictions. We assume the entire ejecta mass  to be in the range
of $10^{-5}$  to $10^{-6}$ $M_{\odot}$ which is fairly representative  for the mass of novae ejecta. Carbon can be  assumed to comprise
about 10 percent of this mass as  indicated by model calculations for elemental abundances (Jose and Hernanz, 1998) in CO novae like V2615 Oph. Thus, in the initial saturated phase when $M_{CO}$ is found to be 3x$10^{-8}$ $M_{\odot}$, the CO:C ratio is determined to lie in the range between $10^{-1}$ and $10^{-2}$. This would clearly rule out the PR models in which the initial abundances of carbon are less than oxygen (model A) and be consistent with model B (carbon rich) where a CO:C ratio of $\sim$ $10^{-1}$ is expected. At later stages, after the destruction of CO, $M_{CO}$ is difficult to estimate precisely but it is certainly lower than one-tenth of its  value during saturation. This suggests that CO:C is in the range $10^{-2}$ to $10^{-3}$ and likely to be even lower.
This is again reasonably consistent with the PR results which find that in most models, regardless of whether the ejecta is O or C rich the CO:C ratio decreases with time and is less than $10^{-3}$ at a  time greater than 50 days.

\section{ Summary} Near-infrared spectroscopy and photometry of the dust forming nova
V2615 Oph are presented. The key observational result is the detection of first overtone
CO emission in this nova. Modeling of the data indicates an initial phase when the
CO is saturated and whose duration $t_s$  is constrained in the range 7 $\le$ $t_s$ $\le$
16 days. During this phase,
the gas temperature and mass are found to be fairly constant  in the  range of  4000 - 4300K and
2.75x$10^{-8}$  to  3.25x$10^{-8}$ $M_{\odot}$ respectively. A ratio of $^{12}$C/$^{13}$C
 $\ge$ 2 is inferred. Following the saturated phase, the CO is found to  be destroyed fairly fast.
The observed timescales involved in the evolution of the CO emission and the estimated
CO:C ratio
are compared with model predictions and found to be in good agreement.

\section*{Acknowledgments}

The research work at Physical Research Laboratory  is funded by the Department
of Space, Government of India.  We are thankful for the online availability of
the Kurucz and NIST atomic linelists and also the optical photometric data from
the AAVSO and AFOEV databases. We wish to thank the anonymous referee for his suggestions and comments which greatly helped in improving the presentation of the paper.

\end{document}